\documentstyle[psfig]{mn}
\newif\ifAMStwofonts

\newcommand {\ie} {{\it i.e.,} }
\newcommand {\eg} {{\it e.g.} }
\newcommand {\etal} {{\it et al.} }

\def\a4{\hsize 15.cm \vsize 23.cm}
\def\skipline{\vskip 10.1pt}

\def\kms{\thinspace\hbox{$\hbox{km}\thinspace\hbox{s}^{-1}$}}

\def\cm3{\thinspace\hbox{$\hbox{cm}^{3}$}}
\def\cm2{\thinspace\hbox{$\hbox{cm}^{2}$}}

\def\hb{\hbox{$\hbox{H}_\beta$}}

\def\dyncm2{\thinspace\hbox{$\hbox{dyn}\thinspace\hbox{cm}^{-2}$}}
\def\deg{\hbox{$^\circ$}}

\def\C2{$\lambda\lambda$\thinspace\hbox{8542,8662}~\AA}

\medskip
\def\t10#1{$\times 10^{#1}$}            
\def\10#1{$ 10^{#1}$}                   
\def\sol{$_\odot$ }                     
\def\pcm3{cm$^{-3}$ }                   
\def\cm2{cm$^{-2}$ }                    

\def\p3{$\pi$}                          

\def\0{$_0$ }                           
\def\etal{{\it et al.} }                
\def\si{$\sim$ }                        
\def\ie{{\it i.e.} }                    
\def\eg{{\it e.g.} }                    
\def\GTT{{\ Tenorio-Tagle} }            

\def\A+A#1{{\it Astron}. {\it Astrophys}. {\bf #1}}

\def\etal   {{\sl et\nobreak\ al.\ }}

\newcount\notenumber

\def\note{\advance\notenumber by 1
  \footnote{$~{\the\notenumber}$}}
\newcount\natrefreg

\def\natref{\advance\natrefreg by 1 $~{\the\natrefreg}$}
\newcount\secreg

\newcount\subsecreg
\def\clearsubsecreg{\subsecreg=0}
\def\newsection#1{\skipline\skipline\par\smallbreak\noindent\advance\secreg by1
 \the\secreg . {\bf #1}\nobreak\par\clearsubsecreg}
\newcount\tablereg

\def\nextable{\advance\tablereg by1\table}
\newcount\figreg

\def\fig{Fig.~\the\figreg}
\def\nextfig{\advance\figreg by1\fig}
\def\RC2{{\ninesl Second Reference Catalogue of Bright Gal\-axies\/},
 de~Vaucouleurs, de~Vaucouleurs and Corwin (1976)}
\def\rc2{\RC2}

\def\x10#1{\hbox{$\times\hbox{10}^{#1}$}}

\def\la{Ly$\alpha$}
\def\hb{H$\beta$}

\def\kms                 {km\thinspace s$^{-1}$}

\def\kms{km\thinspace s$^{-1}$}                        

\def\beq{\begin{equation}}                          
\def\eeq{\end{equation}}                              
\def\beqa{\begin{eqnarray}}                         
\def\eeqa{\end{eqnarray}}                             
\def\beqan{\begin{eqnarray*}}                      
\def\eeqan{\end{eqnarray*}}                          

\def\kms{km\thinspace s$^{-1}$}                        

\def\beq{\begin{equation}}                          
\def\eeq{\end{equation}}                              
\def\beqa{\begin{eqnarray}}                         
\def\eeqa{\end{eqnarray}}                             
\def\beqan{\begin{eqnarray*}}                      
\def\eeqan{\end{eqnarray*}}                          

\ifoldfss
  \ifCUPmtlplainloaded \else
    \NewTextAlphabet{textbfit} {cmbxti10} {}
    \NewTextAlphabet{textbfss} {cmssbx10} {}
    \NewMathAlphabet{mathbfit} {cmbxti10} {} 
    \NewMathAlphabet{mathbfss} {cmssbx10} {} 
  \fi
  \ifAMStwofonts
    \ifCUPmtlplainloaded \else
      \NewSymbolFont{upmath} {eurm10}
      \NewSymbolFont{AMSa} {msam10}
      \NewMathSymbol{\upi}     {0}{upmath}{19}
      \NewMathSymbol{\umu}     {0}{upmath}{16}
      \NewMathSymbol{\upartial}{0}{upmath}{40}
      \NewMathSymbol{\leqslant}{3}{AMSa}{36}
      \NewMathSymbol{\geqslant}{3}{AMSa}{3E}

      \let\leq=\leqslant \let\le=\leqslant
      \let\geq=\geqslant \let\ge=\geqslant
    \fi
  \fi
\fi 

\ifnfssone
  \newmathalphabet{\mathit}
  \addtoversion{normal}{\mathit}{cmr}{m}{it}
  \addtoversion{bold}{\mathit}{cmr}{bx}{it}
  \newmathalphabet{\mathbfit} 
  \addtoversion{normal}{\mathbfit}{cmr}{bx}{it}
  \addtoversion{bold}{\mathbfit}{cmr}{bx}{it}
  \newmathalphabet{\mathbfss} 
  \addtoversion{normal}{\mathbfss}{cmss}{bx}{n}
  \addtoversion{bold}{\mathbfss}{cmss}{bx}{n}
  \ifAMStwofonts
    \ifCUPmtlplainloaded \else
      \UseAMStwoboldmath
      \makeatletter
      \new@mathgroup\upmath@group
      \define@mathgroup\mv@normal\upmath@group{eur}{m}{n}
      \define@mathgroup\mv@bold\upmath@group{eur}{b}{n}
      \edef\UPM{\hexnumber\upmath@group}
      \new@mathgroup\amsa@group
      \define@mathgroup\mv@normal\amsa@group{msa}{m}{n}
      \define@mathgroup\mv@bold\amsa@group{msa}{m}{n}
      \edef\AMSa{\hexnumber\amsa@group}
      \makeatother
      \mathchardef\upi="0\UPM19
      \mathchardef\umu="0\UPM16
      \mathchardef\upartial="0\UPM40
      \mathchardef\leqslant="3\AMSa36
      \mathchardef\geqslant="3\AMSa3E

      \let\leq=\leqslant \let\le=\leqslant
      \let\geq=\geqslant \let\ge=\geqslant
    \fi
  \fi
\fi 

\ifnfsstwo
  \DeclareMathAlphabet{\mathbfit}{OT1}{cmr}{bx}{it}
  \SetMathAlphabet\mathbfit{bold}{OT1}{cmr}{bx}{it}
  \DeclareMathAlphabet{\mathbfss}{OT1}{cmss}{bx}{n}
  \SetMathAlphabet\mathbfss{bold}{OT1}{cmss}{bx}{n}
  \ifAMStwofonts
    \ifCUPmtlplainloaded \else
      \DeclareSymbolFont{UPM}{U}{eur}{m}{n}
      \SetSymbolFont{UPM}{bold}{U}{eur}{b}{n}
      \DeclareSymbolFont{AMSa}{U}{msa}{m}{n}
      \DeclareMathSymbol{\upi}{0}{UPM}{"19}
      \DeclareMathSymbol{\umu}{0}{UPM}{"16}
      \DeclareMathSymbol{\upartial}{0}{UPM}{"40}
      \DeclareMathSymbol{\leqslant}{3}{AMSa}{"36}
      \DeclareMathSymbol{\geqslant}{3}{AMSa}{"3E}

      \let\leq=\leqslant \let\le=\leqslant
      \let\geq=\geqslant \let\ge=\geqslant
    \fi
  \fi
\fi 

\ifCUPmtlplainloaded \else
  \ifAMStwofonts \else 
    \def\upi{\pi}
    \def\umu{\mu}
    \def\upartial{\partial}
  \fi
\fi

\title[Superbubbles in star-forming galaxies]
 {The evolution of superbubbles and the detection of Ly$\bmath{\alpha}$ 
in star-forming galaxies}
\author[G. Tenorio-Tagle et al.]
       {Guillermo Tenorio-Tagle,$^1$\thanks{e-mail: gtt@inaoep.mx} 
Sergey A. Silich,$^2$  Daniel Kunth,$^3$ \newauthor 
Elena Terlevich$^1$ and Roberto Terlevich$^4$\thanks
{Visiting Professor at INAOE.}\\
       $^1$   Instituto Nacional de Astrof\'\i sica Optica y Electr\'onica, 
AP 51, 72000 Puebla, M\'exico.\\
        $^2$   Main Astronomical Observatory National Academy 
of Sciences of Ukraine, 252650, Kiev, Golosiiv, Ukraine. \\
        $^3$ Institut d'Astrophysique de Paris, 98bis Bld Arago, F-75014 
Paris, France.\\ 
        $^4$ Institute of Astronomy, University of Cambridge, Madingley 
Road, Cambridge CB3 0HA, UK.
}
\date{Accepted ...
      Received ...;
      in original form ...}

\pagerange{\pageref{firstpage}--\pageref{lastpage}}
\pubyear{1994}

\begin{document}

\maketitle

\label{firstpage}

\begin{abstract}
The detection of \la\ emission in star-forming galaxies  in different 
shapes and intensities (always smaller than 
predicted for case B recombination)  has puzzled
the astronomical community for more than a decade.
Here we use two dimensional 
calculations to follow the evolution of superbubbles and of the 
H\,{\sevensize II} regions 
generated by the output of UV photons from massive stars. We
show the impact caused by massive star formation in the ISM
of different galaxies and we  look at the conditions required to detect  
\la\ emission  from a nuclear H\,{\sevensize II} region, 
and the variety of profiles that may be expected as a function of time. 

\end{abstract}

\begin{keywords}
H\,{\sevensize II} regions -- galaxies: starburst --galaxies: superbubbles -- 
ISM: structure -- ISM: evolution -- methods: numerical.
\end{keywords}

\section{Introduction}

It has been conjectured that primeval galaxies at large redshift would be 
easily detected from their \la\ emission (Partridge and Peebles 1967; 
Meier 1976). Ultraviolet observations of nearby starburst galaxies however, 
have 
revealed a  much weaker \la\ emission than predicted by simple models 
of galaxy formation (Meier \& Terlevich 1981, Hartmann \etal 1984, 1988,
Deharveng \etal 1986,  
Terlevich \etal 1993, Kunth \etal 1997, 1998, 1999). 
Star-forming galaxies have also been 
recognized at large redshifts and many of them show weak \la\ emission or 
none at all (Steidel \etal 1996, Lowenthal \etal 1997).  The reason for the
weakness of the \la\ emission has been the subject of a series of debates. 
It was early realized that pure extinction by dust would be unable to
explain the low observed \la /\hb\ although this was disputed by Calzetti \&
Kinney (1992) who tentatively proposed that proper extinction laws would
correctly match the predicted recombination value. Valls-Gabaud (1993) 
on the other hand suggested that ageing  starbursts  have reduced \la\ 
equivalent widths affected by strong underlying  stellar atmospheric 
absorptions.  Early IUE data has provided evidence for a correlation of \la\ 
emission and the metallicity of the gas as expected if \la\ is destroyed by 
dust absorption (Meier \& Terlevich, 1981, Hartmann \etal 1988, Terlevich 
\etal 1993). Dust absorption can be particularly effective if multiple 
scattering in neutral hydrogen selectively increases the path length of these 
photons through the dusty regions (Auer, 1968). Moreover high density 
H\,{\sevensize I} regions can also backscatter photons into the 
H\,{\sevensize II} region in which even a small amount of dust mixed within
the ionized gas could be an important agent of destruction. Detailed 
calculations performed by Chen \& Neufeld (1994) show how these effects can 
lead to \la\ absorptions  even for galaxies in which the unattenuated spectrum
would show a strong \la\ emission line.

Then, as also shown by Charlot \& Fall (1993) 
the structure of the interstellar medium (porosity and multi-phase structure) 
is most probably an important factor. As a result the  
\la\ emission {\it vs}  metallicity correlation should show a large scatter, as 
evidenced in the recent compilation of 21 local low metallicity 
starburst galaxy spectra from the IUE archives (Giavalisco \etal 1996).

New HST data on ten star-forming galaxies 
(Lequeux \etal 1995; Kunth \etal 1997, 1998; 
Thuan \& Izotov 1997) has allowed the further 
realization that the velocity structure in the interstellar 
medium plays also a key role in the transfer and escape of \la\
photons (Kunth \etal 1997). 
Three types of observed line profiles have been identified: pure \la\ 
emission;  broad damped  \la\ absorption  centered at the 
wavelength corresponding to the redshift of the H\,{\sevensize II} emitting gas; and \la\
emission with blue shifted absorption features, leading in some cases 
to P Cygni profiles.
As noted by Kunth \etal (1998),  \la\ emission with deep absorption 
troughs at the 
blue side of the \la\ profiles, 
evidence of a wide velocity field, has been detected in four of 
the observed galaxies. 
Interstellar absorption lines (O\,{\sevensize I}, Si\,{\sevensize II}) are
also significantly blueshifted with respect to the H\,{\sevensize II} gas.

The velocity and density structure of the neutral gas along the line of sight 
rather than the 
abundance of dust particles alone, seem to be the determining factor 
for the escape of 
\la\ photons in these objects. Lequeux \etal (1995) in the
analysis of the HST data of Haro 2, proposed that 
if most of the neutral gas is 
outflowing from the ionized region,  perhaps in a galactic wind 
resulting from the intense star 
formation activity, then \la\ photons will escape (partially) unaffected.
If the H\,{\sevensize I} is static with respect to H\,{\sevensize II}, the 
destroyed \la\ photons are those emitted by the H\,{\sevensize II} region; 
otherwise, they would correspond to the stellar continuum at wavelengths 
close to \la .
Nine out of the ten galaxies show broad \la\ absorption, with a very narrow derived 
range of logarithmic column densities (19.7 to 21.5 cm$^{-2}$). 

Being aware of the complications intrinsic to the transport of \la\
lines, we postulate here a simple scenario that accounts  
for the variety of \la\ line detections in starburst galaxies.
The scenario is supported by two dimensional calculations of multi-supernova
remnants evolving in gas-rich disk galaxies covering the range between dwarf
and massive galaxies such as the Milky Way (see Sections 2 and 3). The 
calculations account for the photoionization produced by a massive 
starburst, and its impact on the surrounding gas as a function of time. 
Our models are based on the synthetic properties of starbursts, as 
derived by Mas-Hesse \& Kunth (1991) and Leitherer \& Heckman (1995); 
and consider a mechanical energy injection rate  ($E_0$ 
\si  10$^{38} -- 10^{42}$ erg s$^{-1}$)
which over several 10$^7$ yr is equivalent to some tens of thousand of 
supernova explosions, and the corresponding input of ultraviolet photons
generated by the massive stars in the considered 
cluster. The continuous mechanical energy deposition generates a 
remnant with an outer shock that sweeps the surrounding gas into a thin shell,
while the evacuated cavity is filled with the hot ejecta; efficiently 
thermalized at the inner shock. The remnant at first  decelerates rather 
rapidly,
although its  overall expansion allows it to depart from spherical 
symmetry, given  
the lower densities away from the galaxy plane. The latter eventually 
becomes insufficient to cause any further deceleration and the remnant blows 
out (see \GTT \& Bodenheimer 1988; and Silich \& \GTT 1998, hereafter
referred to as Paper I) allowing for the venting of the hot matter into the
galaxy halo.
As shown in Paper I, blowout occurs quite early in the evolution, 
as the superbubbles reach dimensions \si 1 kpc. This leads to 
shell fragmentation and to the venting of the processed matter into the 
extended low density halo where the outer shock forms a new shell of swept 
up matter. This in the case of dwarf galaxies, initially expands
with speeds that well exceed the local escape speed of the galaxy. However, 
its motion into the gaseous halo causes a continuous deceleration lowering 
its velocity to values well below the escape speed of the galaxy.
These facts are in excellent agreement with detailed observations of 
dwarf and irregular galaxies which show giant complete remnants expanding 
at a pace comparable or slower that the escape velocity of their parent galaxy 
(see Marlowe \etal 1995, Martin 1998). Here we consider the fact that blowout 
must  also allow for the escape of a fraction of the UV radiation produced by 
the massive members of the starburst, causing an extended H\,{\sevensize II} 
region into the galaxy halo.
The latter is here also held responsible for the escape of UV and  \la\ photons
from the galaxy. We follow the time evolution of the photoionized region 
(Section 3) accounting for the drop in stellar luminosity caused by the ageing 
of the cluster, as well as for the range of recombination times that arises 
from the different densities that characterize the galaxy haloes and the 
decelerating shells. The calculations allow us to predict (Section 4) a 
variety of \la\ profiles that are to be expected  from these sources, as a 
function of time and inclination angle; and a comparison with the existing 
observations is included in the Discussion section.

\section[]{The evolution of Superbubbles \\* in gas-rich disk galaxies}

\subsection{The ISM density distribution}

We approximate our galaxies with the prescription 
for the stellar, dark matter, and gaseous components
as given in Paper I and in Li \& Ikeuchi (1992).
The gas density distribution  allows for
two isothermal components, related to 
the central dense molecular core (with a temperature of 100 K$^{\circ}$), and  
a low density neutral gas halo (with temperature 1000 K$^{\circ}$). 

\begin{equation}
      \label{a.1}
\rho_{g} = \rho_{core} + <\rho_{halo}>
\end{equation}
Both components are supported in a quasi-equilibrium state by rotation 
and random gas motions with an effective gas pressure 
\begin{equation}
      \label{a.2}
P_{ext} = \frac{1}{3} (\rho_{core} C_{core}^2 +
          <\rho_{halo}> C_{halo}^2),
\end{equation}
where 
$C_{core}$ is the  velocity dispersion of the 
compact dense core, and
$C_{halo}$  that of the extended halo component. 

\begin{figure}
\psfig{figure=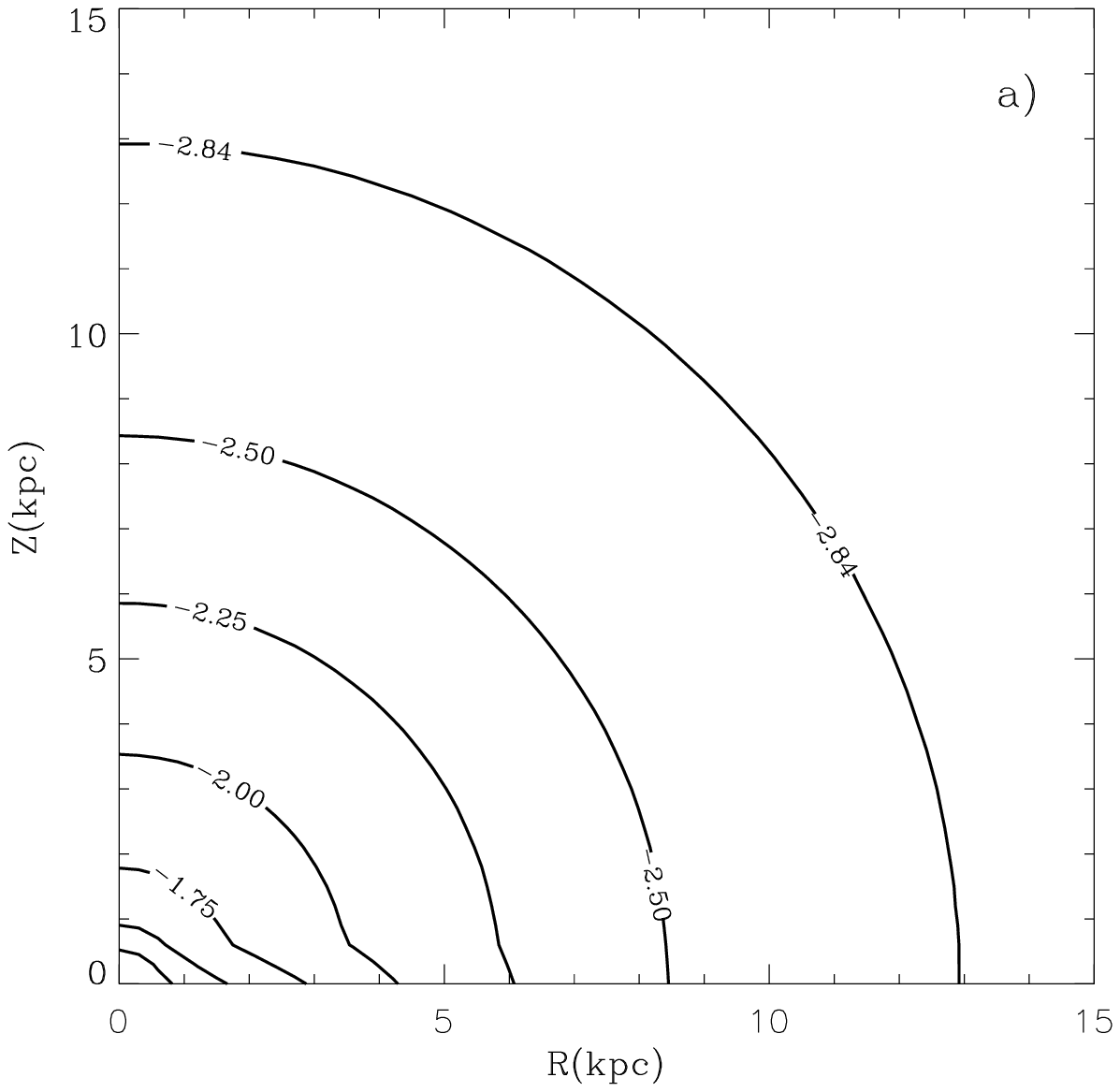,height=7cm,width=9cm,angle=0}
\psfig{figure=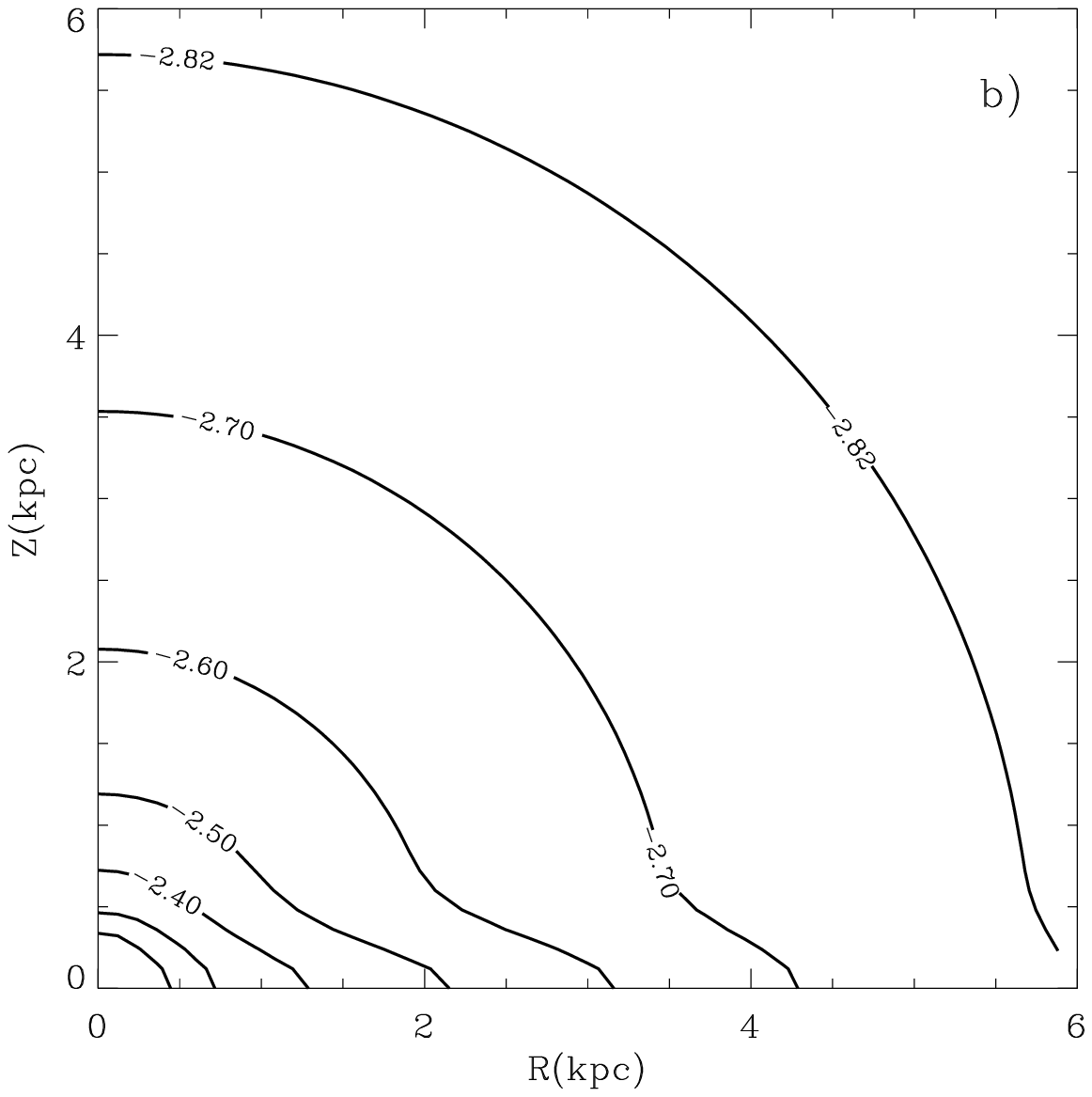,height=7cm,width=9cm,angle=0}
\psfig{figure=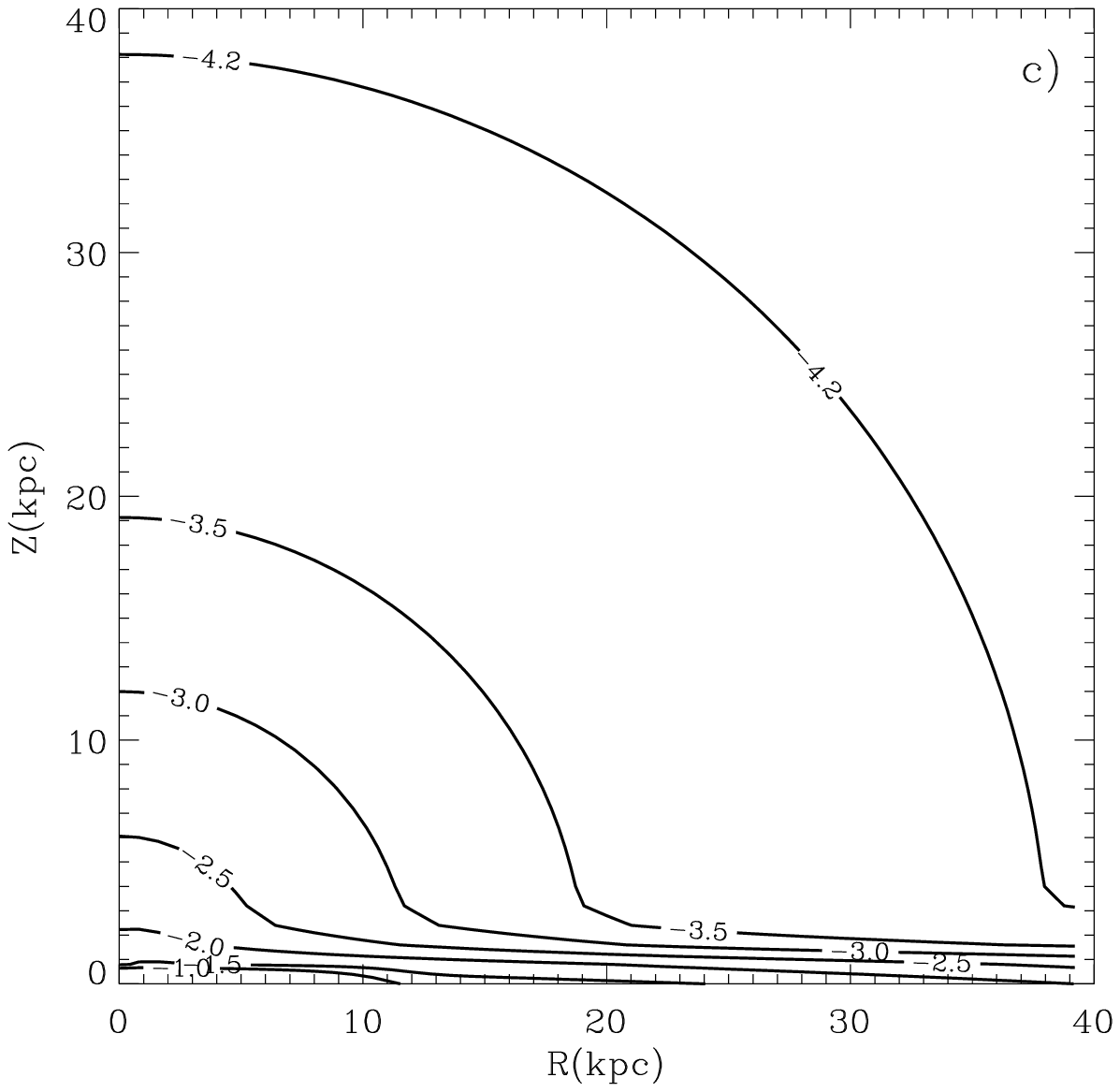,height=7cm,width=9cm,angle=0}
 \caption{ISM gas distribution. Panels a, b, and c display the initial gas
distribution of the $10^{9}$,
$5 \times 10^{7}$, and 10$^{10}$ M\sol for models A, B and C, 
respectively (see Table 1).}
\label{fig1}
\end{figure}

Typical initial density distributions, for several of the cases here 
considered (see Table 1), are shown in Fig.~\ref{fig1}a-c. In all cases the ISM mass 
amounts to 10 percent of the total mass of the galaxy and ranges from 5 
$\times 10^7$ M\sol
in the case of the smallest dwarf galaxies, to $10^{10}$ M\sol for the more
massive disk-like galaxies considered here.

\begin{figure}
\psfig{figure=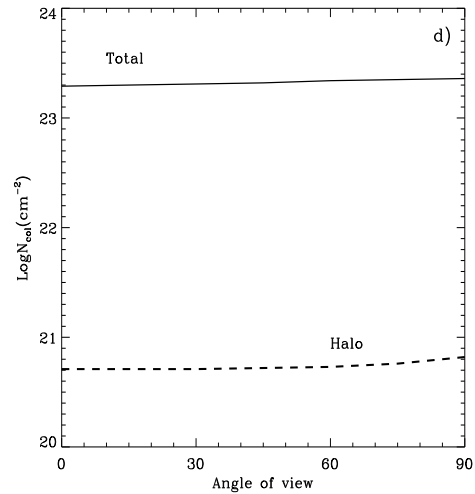,height=7cm,width=9cm,angle=0}
 \caption{Total and halo initial column density 
(cm$^{-2}$) for model A.}
\label{fig2}
\end{figure}

The initial column densities
vary between $10^{20}$ and $10^{23}$ cm$^{-2}$ for different models.
Note however, that only a minor part of this value comes from the extended low
density halo. Most of the undisturbed gas column density is
(as shown in  Fig.~\ref{fig2}) associated with
the central molecular core. 

\subsection{The starburst mechanical energy and \\* UV radiation input rates} 

The mechanical luminosity from massive starbursts ($M_{stars}$ $\geq$ 10$^6$ 
M\sol) is known to lead to a rapidly evolving superbubble able to blowout from 
the gaseous galactic disk configuration into the extended H\,{\sevensize I} 
halo, as $E_0$ exceeds the energy input rate required for the remnant to 
reach a typical disk scale-height with a supersonic velocity. As shown 
by Koo \& McKee (1992) if $E_0$ exceeds the threshold luminosity 
$L_b$ = 7.2 $ \times 10^{36} P_4 H_{100}^2 a_{s,10}$  erg s$^{-1}$
(where $P_4$ = disk pressure in units of $(10^4k)$ cm$^{-3}$ K, $H_{100}$ is 
the disk scale height in units of 100 pc and $a_{s,10}$ is the disk sound 
speed in units of 10 \kms) the superbubble will blowout; thus massive 
starbursts are expected to lead to the superbubble blowout phenomena even in 
massive galaxies such as the Milky Way. One can then predict  the venting of 
the hot superbubble interior gas through the Rayleigh-Taylor fragmented 
shell, into the extended H\,{\sevensize I} halo where it would push once again 
the outer shock allowing it to build a new shell of swept-up halo matter. 
A coeval starburst with a total mass larger than several
$10^5 -- 10^6$M\sol, also produces an overwhelming ionizing photon flux 
($F_{UV} \geq 10^{52} $ photons s$^{-1}$) and then blowout
must also allow the leakage of at least a fraction of
the UV photons emitted by the starburst.
These photons are here shown to establish an ionized conical 
H\,{\sevensize II} region with its apex at the starburst. 

It is stellar evolution and the physics of H\,{\sevensize II} regions what 
defines the time-scales relevant to the evolution, detection and impact that 
massive bursts of star formation may have on the ISM. 
Once the mass of a coeval starburst and its IMF are fixed, the integrated 
mechanical energy power of winds and supernovae will remain almost constant 
for up to 40 Myr; the life-time
of an 8 M\sol\ star. On the other hand, the UV photon production rate
will remain constant until  the most massive members of the cluster 
begin to move away from the main sequence (3 -- 4 Myr) to shortly end-up 
as supernovae. From then onwards the starburst flux of UV photons decays 
rapidly  (as $t^{-5}$; see Beltrametti \etal 1982) and thus after 10Myr, once 
the 20 M\sol\ stars have also exploded as supernovae, the H\,{\sevensize II} 
region gas begins to recombine. Detectability of the ionized gas is thus 
restricted in the optical and radio recombination lines to 10 Myr, while the 
impact of winds and supernovae is a much longer lasting event. One can 
trace in X-rays the thermalized ejected matter either within a superbubble 
or in a galactic 
wind emanating from a nuclear starburst, until it cools either adiabatically 
or by radiation ($t_{cool} = 3kT/(2n\Lambda)$ \si 10$^8$yr; where $n$ and
$T$ are the average density and temperature in the bubble, $k$ is the 
Boltzmann constant and $\Lambda$ the corresponding cooling rate). 
The remnants of the   
accelerated ISM can also be detected during a time scale (\si 10$^8$ yr)
that well exceeds the energy deposition time  until they slow down to
speeds comparable to the random speed of motions of the ISM in their host 
galaxies (\si 10 \kms). Shells, loops and funnels in a variety of 
sizes and shapes have been detected in H\,{\sevensize I} and in 
ionized transitions in many starbursts, dwarf, and irregular galaxies
(\eg Marlowe \etal 1995, Hunter \& Gallagher 1997 and  Martin 1998) 

Our models are restricted to realistic mechanical energy input rates 
and to the corresponding number of UV photons emitted by the massive
members in the stellar clusters. Our standard model A assumes a $10^6$ M\sol
instantaneous starburst with Z = 0.25 Z\sol metallicity, and a Salpeter IMF 
with a range of stellar masses between 1 and 100M\sol, causing a constant
mechanical luminosity of $3.5
\times 10^{40}$ erg s$^{-1}$ for up to 40 Myr. This is consistent with 
both the synthetic properties of starbursts
(see Mas-Hesse \& Kunth 1991 and Leitherer \& Heckman 1995) and with 
the values derived by Marlowe \etal (1995) from observations
of intermediate mass galaxies with a nuclear starburst.
The corresponding number of UV photons produced by the massive stars in the 
cluster ($N_{sb}=10^{53}$ photons s$^{-1}$) was regulated to 
remain constant for up to 4Myr, and then, following stellar evolution models, 
to drop as $(t/4Myr)^{-5}$. A fraction of the UV photons produced by the 
starburst were assumed to be used
within a starburst region, and therefore only
$N_{UV} = \varepsilon N_{sb}$ could escape out of the central region and
ionizes the superbubble outer shell, and the extended unperturbed halo. 
Values of $\varepsilon$ between 0.1 and 0.7 were used in our grid of models. 
The mechanical energy input rate and the UV flux produced by the starburst
were then linearly scaled with the mass of the host galaxy. This led to 
values of 1.4 $\times 10^{39}$ erg s$^{-1}$ and 4 $\times 10^{51}$ UV 
photons s$^{-1}$, for the smallest systems, and  3.5 $\times 10^{42}$ erg 
s$^{-1}$ and N$_{sb}$ = 10$^{55}$ UV photons s$^{-1}$ for the giant nuclear 
starbursts
in the massive disk galaxies here considered. A second calculation
of the massive galaxy powered by a constant star formation rate instead of a 
coeval starburst is also shown (case C2). In this case the total 
mechanical energy power and UV radiation flux of case C1 were assumed to 
remain constant and spread over the 10$^8$ yr of the considered evolution. 
The main input parameters that define our galaxies -- starburst strength and 
boundary conditions -- are collected in Table 1 which includes in columns 
1 -- 3 the total mass, the total ISM mass and the central densest ISM mass. 
Columns 4 and 5 list the radius of the dark matter distribution and the radius 
of the galaxy's ISM (see Paper I), $C_{halo}$ and $C_{core}$ are the assumed 
values for the 
velocity dispersion of the dense core and the extended halo components; $n_0$
is the central number density. Columns 9 and 10 list the mechanical power of 
the assumed starburst and its total UV output. Column 11 lists the assumed 
metallicity of the ISM and column 12 the starburst life-time.  
Details of the computational approach used to derive the impact of
the ionizing radiation on the evolving giant remnants 
and the extended halos are given in section 3.

  \begin{table*}
 \centering
 \begin{minipage}{160mm}
 \caption{Model parameters.}
\begin{tabular}{lccccrrrrcccr} 
& $ M_{tot} $ & $ M_{gas} $ &$ M_{core} $ &  $ R_B $ & 
$ R_G $ & $ C_{halo} $ & $C_{core}$  &  $ n_{0} $ &  L  & N$_{sb}$ &
  Z/Z$_{\odot}$ & t$_{burst}$ \\
N & $10^{10} M_{\odot}$ & $10^9 M_{\odot}$ & $10^7 M_{\odot}$ & kpc 
& kpc &km/s$^{-1}$ & km/s$^{-1}$ &cm$^{-3}$ 
&$10^{40}$erg s$^{-1}$& $10^{52}$s$^{-1}$ & & Myrs\\[10pt]
A1 & 1.0 & 1.0 & 5.0 & 10 & 13.6 & 80 &20 &1.4 $\times$10$^3$& 3.5 &
10.0 & 0.3 & 40 \\
A2 & 1.0 & 1.0 & 5.0 & 10 & 13.6 & 80 &20 &1.4 $\times$10$^3$ & 3.5 &
10.0 & 1.0 & 40 \\
A3 & 1.0 & 1.0 & 5.0 & 10 & 13.6 & 80 &20 &1.4 $\times$10$^3$ & 3.5 &
10.0 & 0.1 & 40 \\
B1 & 0.05& 0.05& 0.1 & 10 & 5.9 & 25  & 5 & 18.6 & 0.14 & 0.4 & 0.3 & 40 \\
C1 & 10.0 & 10.0 &500.0 & 20 & 39.6 & 150 &20 & 17.0 & 350 & 10$^3$ & 0.3&
     40  \\
C2 & 10.0 & 10.0 &500.0 & 20 & 39.6 & 150 &20 & 17.0 & 140 & 4$\times$10$^2$
& 0.3 & 100 \\
\end{tabular}
\end{minipage}
\end{table*}

\section{The Numerical model}

Our hydrodynamical models describe in detail (see Paper I) the shell 
morphology, surface density and expansion velocity, as a function of time.
The shock velocity, and post-shock densities and temperatures then 
follow from the Rankine-Hugoniot jump conditions.
The Mach number $M_s = D_s / C_{ism}$ depends on the ionization state 
of the ISM through the changes in the ambient gas sound velocity 
\begin{equation}
      \label{a.4}
C_{ism} = \left(\frac{\gamma k T_{eff}}{\mu}\right)^{1/2},
\end{equation}
where the mean mass per particle is $\mu_n = 14/11m_H$ 
in the neutral region, and  $\mu_i = 14/23m_H$ within the ionized zone. The 
effective gas temperature was taken to be
\begin{equation}
      \label{a.5}
T_{eff} = \frac{P_{th}}{k n_g},
\end{equation}
where $P_{th}$ is the thermal pressure and k the
Boltzmann's constant. We then calculate the post-shock gas 
adiabatic - radiative transition with a tabulated 
cooling function (Gaetz and Salpeter, 1983). 

After blowout, and the disruption of the shell via Rayleigh-Taylor
instabilities, the hot gas streams out of the cavity with a characteristic 
velocity of the order of it's own speed of sound. Therefore we allow the shell
to lose mass up to the end of acceleration, when the
expansion velocity equals the sound velocity at the bubble center. 
After blowout the halo gas begins to be  collected by a rapid (adiabatic)
shock, and a new thin radiative shell forms when the  characteristic 
cooling time exceeds the time interval between blowout and the current 
evolution time $t$. 

\subsection{Shell structure}

The structure of the shell of swept-up matter was assumed to be able to 
develop   
two concentric zones, and the inner shell surface was derived from the 
numerically calculated positions. The full  
thickness of the shell at every Lagrangian mesh is 
\begin{equation}
      \label{a.6}
l_s = \frac{\sigma}{n_s \mu_n},
\end{equation}
where $\sigma$ is the shell surface density, and $\mu_n$ and $n_s$ are the mean
mass per particle and post-shock gas density, respectively. We then calculate 
the number of photons required to ionize an equivalent spherical shell with
the same Lagrangian segment radius R, surface density, and shell number 
density:
\begin{equation}
      \label{a.7}
N_c = \frac{4 \pi}{3} [(R+l_s)^3 - R^3] n_s^2 \alpha_B,
\end{equation}
where $\alpha_B=2.59 \times 10^{-13}$ cm$^3$  s$^{-1}$  is the H
recombination coefficient to all levels but the 
ground state (Osterbrock, 1989). Taking into account the angle $\alpha_i$ 
between the vector normal to the shell surface and the radius-vector
$R_i$ of a particular gridpoint, one can define the ratio
\begin{equation}
      \label{a.8}
\beta = \frac{N_s}{N_c} ,
\end{equation}
where $N_s = N_{UV} \cos \alpha_i$. This allows one to distinguish between 
three obvious cases along each radial direction: a fully ionized shell
($\beta \ge 1$), a neutral shell ($\beta = 0$), and a partially ionized
shell ($0 < \beta < 1$), where all UV photons available are trapped within
the shell. In the later case the condition 
\begin{equation}
      \label{a.9}
N_s = \frac{4 \pi}{3} [(R+l_i)^3 - R^3] n_s^2 \alpha_B,
\end{equation}
where $l_i$ is the thickness of the ionized  shell layer should be
fulfilled. Combining (\ref{a.9}) with equations (\ref{a.7}) and (\ref{a.8})
gives the ionized skin thickness:
\begin{equation}
      \label{a.10}
l_i = \left[\beta[(R+l_s)^3 - R^3] + R^3\right]^{1/3} - R .
\end{equation}
The width of the neutral skin then follows from the difference between
the full shell thickness (\ref{a.6}) and the thickness of the ionized layer  
(\ref{a.10})
\begin{equation}
      \label{a.11}
l_n = R + l_s - \left[\beta[(R+l_s)^3 - R^3] + R^3\right]^{1/3}.
\end{equation}

\subsection{The ionization of the ISM}

Our simplified approach considers the ionization along a limited number of 
lines of sight (typically along every 15$^{\circ}$ from the axis of symmetry)
discarding the shell Doppler shift for photon absorption and the UV photons 
spectral distribution. Photons moving along lines of sight
that inpinch on neutral sections of the shell, are then used up in its
ionization. If the shell segment is in its adiabatic stage, it is then
transparent to the ionizing radiation for as long as the post-shock
temperature exceeds 10$^4$ K. 

The UV photons that escape the shell cause then the ionization of the halo.
Following an ionization equilibrium equation and the on-the-spot 
approximation, without dust absorption (Osterbrock, 1989), the ionization of 
each shell section and the corresponding external interstellar gas is
considered separately: 
\begin{eqnarray}
      \label{a.12}
     & & \hspace{-0.5cm}\nonumber
\Delta(R) = N_{UV} - 
\\[0.2cm]
      & & \hspace{-0.2cm}
4 \pi \left(\int_{R_1}^{R_2} n^2(r) \alpha_B r^2 
{\rm d}r + \int_{R_2}^R n^2(r) \alpha_B r^2 {\rm d}r \right), 
\end{eqnarray}
where $R_1$ and $R_2$ are the inner and outer shell radii respectively. The 
shell 
densities at every cross-point are linearly interpolated from the values
at the nearest Lagrangian points. It was also assumed that within the shell
the density varies linearly along lines of sight. Then the first integral 
($I_1$) in equation (\ref{a.12}) can be calculated analytically, and reads  
\begin{eqnarray}
      \label{a.13}
     & & \hspace{-0.5cm}\nonumber
I_1 = \frac{R_2^3 - R_1^3}{3} \left[n_1^2 + (n_2 - n_1)^2 \frac{R_1^2}
      {l^2} - 2 n_1 (n_2 - n_1) \frac{R_1}{l} \right] +
\\[0.2cm]
      & & \hspace{-0.2cm} \nonumber
      \frac{(n_2 - n_1)}{2}\left[n_1 - (n_2 - n_1) \frac{R_1}{l} \right]
      \frac{(R_2^4 - R_1^4)}{l} + 
\\[0.2cm]
      & & \hspace{-0.2cm} 
      \frac{(n_2 - n_1)^2}{5}
       \frac{(R_2^5 - R_1^5)}{l^2}, 
\end{eqnarray}
where $l = R_2 - R_1$ is the shell thickness along a line of sight. The second
integral in equation (\ref{a.12}) is calculated numerically.

The value of $\Delta$ is first calculated at the galaxy boundary
($R$ = $R_G$). Positive $\Delta$ values represent the escape of UV photons 
from the galaxy. On the other hand, if $\Delta$ is negative, the number of
photons is not sufficient to ionize all the halo, and the radius of the 
H\,{\sevensize II} region is then determined 
by an iteration procedure to lie between the shock 
front and the galaxy outer boundary, wherever the value of $\Delta$ 
becomes equal to zero. 

After 3.5 -- 4 Myr, the number of UV photons produced by the starburst rapidly
drops. Also, between 4 -- 13 Myr (Models A and B) the leading
shock wave moving in the 
halo becomes radiative and the density within the shell drastically increases
(as the square of the Mach number), causing a larger recombination rate.
Both proceses reduce the UV flux available to keep the galaxy halo fully 
ionized,
and lead to the trapping of the ionization front within the expanding shell.

\begin{figure}
\psfig{figure=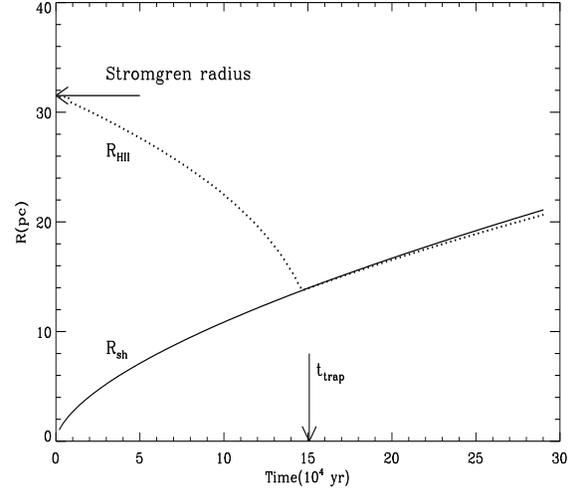,height=7cm,width=8cm,angle=0}
 \caption{The trapping of the ionization front.
The results of a numerical calculation for an
ISM gas number density of 10 particle cm${-3}$, and an 
ionizing photon flux of 10$^{50}$s$^{-1}$.
The arrows indicate the analytic predictions for the Str\"{o}mgren radius 
and the characteristic time when all UV photons are trapped within 
the shell of swept-up matter.}
\label{fig3}
\end{figure}

Our procedure to estimate the trapping of the ionization front,
was checked against the analytic formula for a spherical bubble
expanding in the preionized homogeneous medium (Comeron 1997), and we found
our numerical results in full agreement with analytic predictions 
(see Fig.~\ref{fig3}).

Our procedure also considers the fact that upon photoionization, the 
recombination
time ($\tau_{rec} = 1/n_e \alpha_A \approx 10^5/n_{halo}$ yr) of 
the low density halo gas may exceed 
the characteristic dynamical time of the superbubble. This causes the
ionized halo to become temporarily transparent to the ionizing radiation. 
We thus calculate the characteristic recombination time $\tau_{rec}$ along 
different lines of sight and compare this value with a time interval 
$\tau = t - t_{blowout}$ between a current time $t$ and the time when a shell 
section becomes transparent to the starburst UV radiation, causing the 
ionization of the halo. The number of ionizing
photons which escape the galaxy in any direction follows from the
equation
\begin{equation}
      \label{13}
N_{esc}(\Theta) = N_{UV} - N_{trap}(R_{\tau}),
\end{equation}
where $\Theta$ is the angle between the direction considered and the
symmetry axis. If $\tau_{rec}(R_G) < \tau$, the halo can only be kept
ionized if there is a continuous input of UV photons, and the radius
R$_{\tau}$ equals the halo radius
 (R$_{\tau}$ = R$_G$). Otherwise, if $\tau_{rec}(R_G) \ge \tau$,
recombination within the densest sections of the galaxy halo
will diminish the escape of UV photons. In this case the  radius
R$_{\tau}$  follows
from the condition $\tau_{rec}$(R$_{\tau}$) = $\tau$.

\section{A typical evolutionary sequence}

\begin{figure*}
\centerline{\psfig{figure=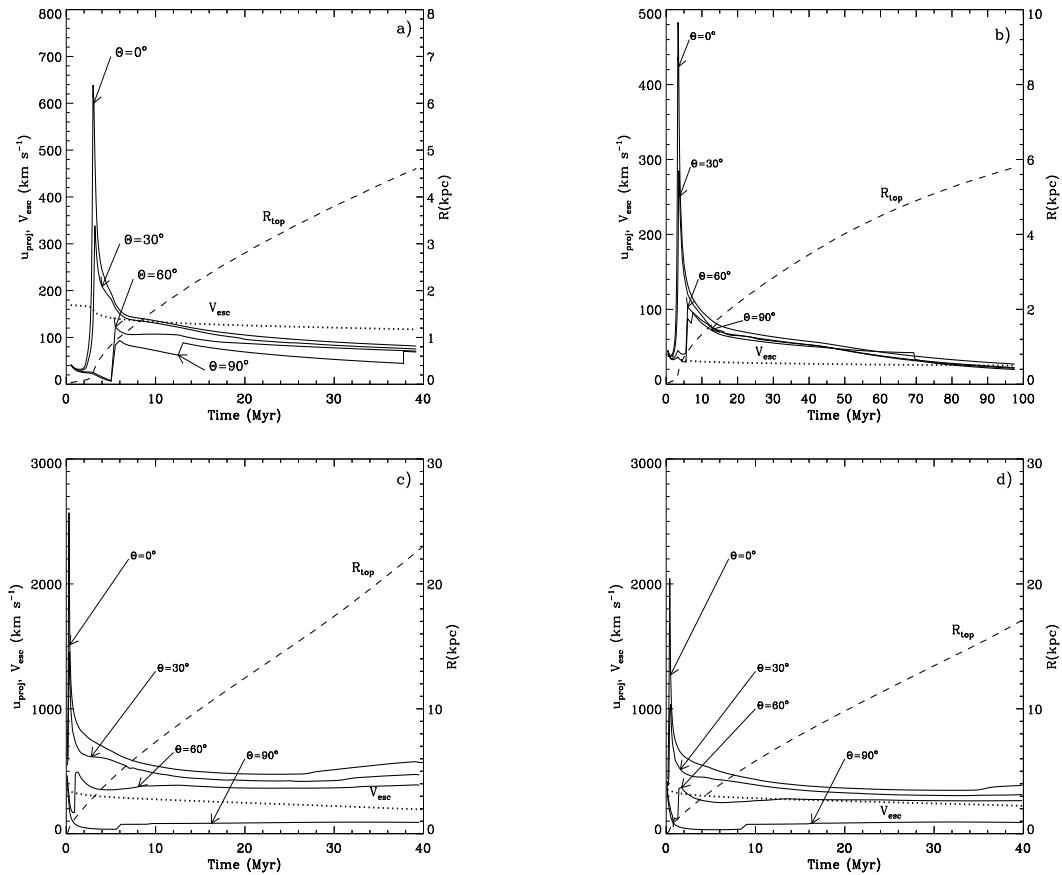,height=12cm,width=16cm,angle=-90}}
 \caption{The evolution of superbubbles.  Panels a-d 
display the results for models A1, B, C1 and C2.
Solid lines represent the
expansion velocities of the superbubbles along different
lines of sight ($\Theta$), and as a function of time. 
These are to be compared with the dotted line which indicates 
the local escape velocity from a galaxy center at a 
distance comparable to 
the maximum dimension of the superbubble (along $\Theta$ = 0),
indicated by the dashed lines.}
\label{fig4}
\end{figure*}

Fig.~\ref{fig4}a shows the evolutionary sequence of our basic model A1. It 
compares the remnant expansion speed along different angles ($\Theta$) 
from the symmetry axis with the galaxy escape speed, and indicates the 
maximum size acquired by the superbubble as a function of time.
The shell expansion velocity shows at first a continuous deceleration,
followed by a rapid acceleration of the pole sections of the remnant 
immediately after blowout and shell disruption. During the acceleration phase 
the shock builds a new shell of swept up halo matter. Afterwards, the 
remnant speed begins to decline even along the direction perpendicular to the 
galaxy plane ($\Theta$ = 0), as the shock gathers more halo matter. In this 
way, the remnant ends up expanding with speeds well below the galaxy escape 
velocity V$_{esc}$ (see also paper I). Nevertheless, after 40 Myrs (the 
starburst lifetime) a huge remnant has developed and extends in the pole 
direction almost 5 kpc out of the starburst region (see Fig.~\ref{fig4}a). 
This late phase in the evolution of the remnant, as stated earlier, 
is unobservable in the optical regime.

\begin{figure*}
\centerline{\psfig{figure=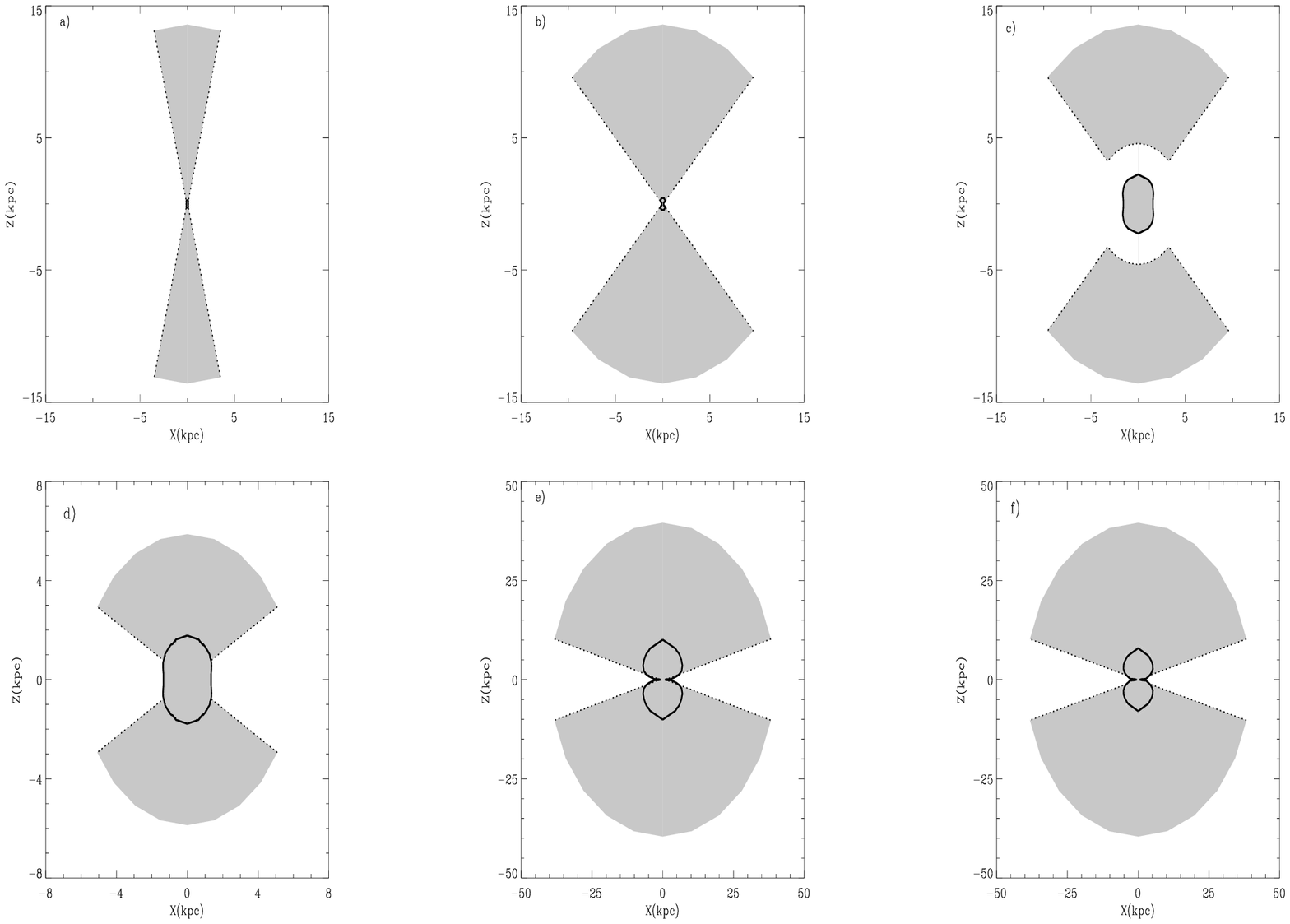,height=12cm,width=16cm,angle=-90}}
 \caption{The extended conical H\,{\sevensize II} regions. Cross-sectional 
views 
along the symmetry axis display the growth of the superbubble (solid contours) 
and the development of the conical H\,{\sevensize II} region (shaded area) in 
model A1 (panels a-c). Note that the H\,{\sevensize II} region extends to the 
edge of the galaxy, reducing in this way the column density of neutral 
material within a wide cone. Panels a-c display the evolution after 3.1, 
3.5 and 15 Myr. Panel c shows the growth of an increasingly wider shell of 
neutral matter generated by recombinations in the halo at late evolutionary 
times. d-f show the results after 15 Myr of evolution for case B, C1, and C2. 
Note that the low densities in the halos of these galaxies have inhibited 
recombinations even after the main H\,{\sevensize II} region life-time.}
\label{fig5}
\end{figure*}

Fig.~\ref{fig5}(a-f) show the growth of the superbubble as well as
the development  of the extended ionized zone and it's evolution  as
a function of time. Before 1.5 Myr the dense shell traps
all UV photons escaping the central starburst region. However,
after blowout and  shell disruption, the  UV photons escaping the central 
H\,{\sevensize II} region produce an initially narrow cone
of ionized interstellar gas around the symmetry axis. The conical 
H\,{\sevensize II} region becomes rapidly broader, and at $t$ \si 
2.5 Myr it reaches it's maximum opening angle. Note that the low density
halo matter has a long recombination time ($t_{rec} = (\alpha n)^{-1}$
\si $10^5$yr/$n$;
where $\alpha$ is the recombination coefficient and $n$ the local density)
and thus upon photoionization it becomes undetectable
both in the optical regime and in the 21 cm line and it also becomes 
transparent to further UV radiation arising from the starburst.
After 5 Myr the superbubble leading 
shock makes a transition from the adiabatic to the radiative regime and 
matter in the swept up shell begins to recombine. 
Given the reduced number of ionizing photons produced by 
the starburst after 4 Myr and the dilution factor 
experienced by the UV radiation to reach the recombining shell,    
rapidly all available photons begin to be  used within the shell
inhibiting  the further leakage 
of UV radiation into the halo and out of the galaxy. Then, after 8 Myr, the 
densest parts of the still unshocked halo begin to recombine. 
Recombination proceeds outwards from the densest 
unshocked halo regions, forming an increasingly wider layer of 
neutral gas ahead of the expanding shell (see Fig.~\ref{fig5}c).

\begin{figure*}
\centerline{\psfig{figure=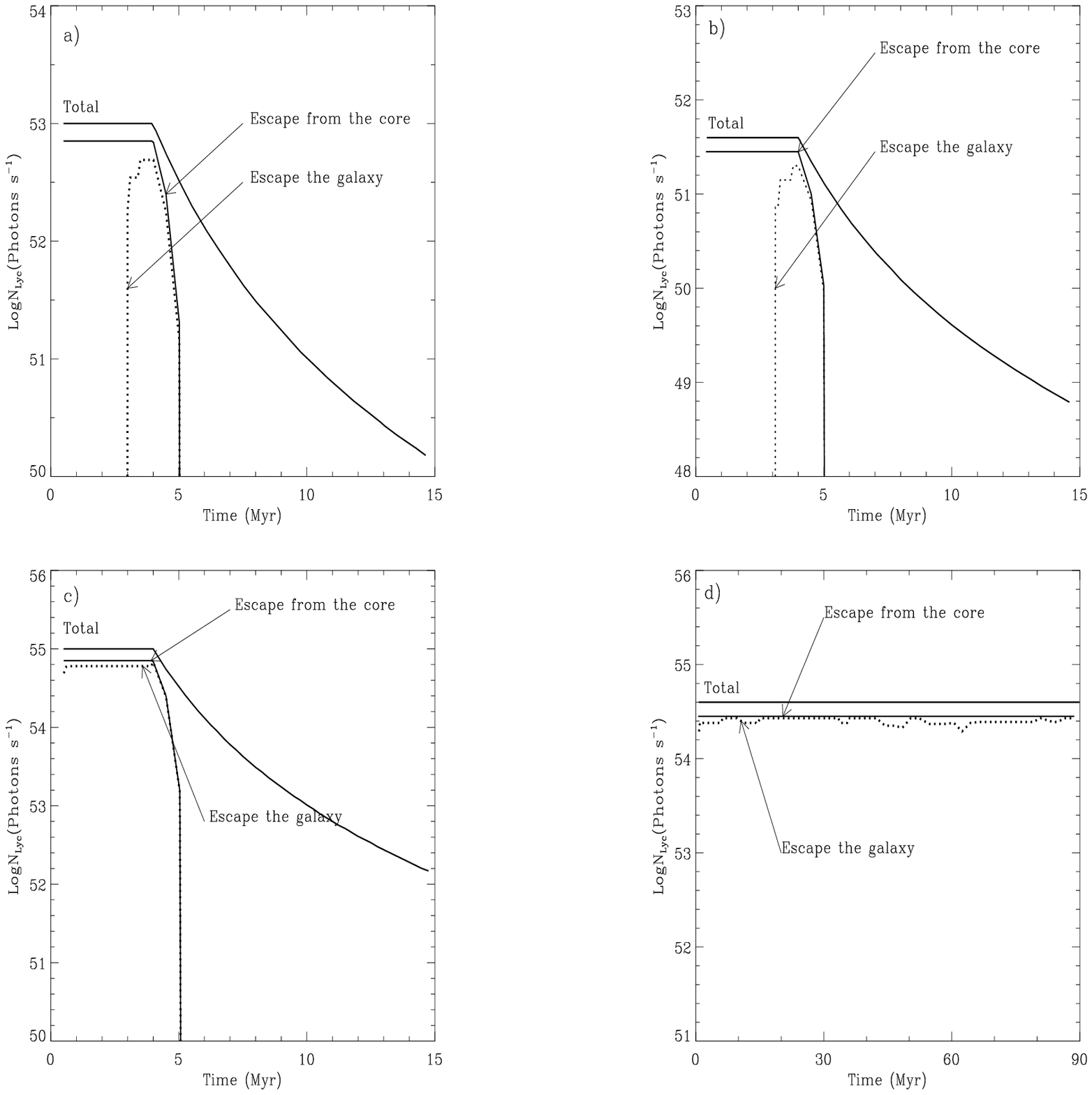,height=12cm,width=16cm,angle=-90}}
 \caption{The number of UV photons that escape from a galaxy in models A, B, 
C1 and C2 are shown in panels a,b,c and d respectively, as a function of time.
The escape of UV radiation does not start until the whole galaxy halo 
is ionized, and in all coeval starburst models (A--C1) it is restricted to 5Myr
by the condition imposed by stellar evolution. Note that given the low 
densities in the galaxy halos and the corresponding recombination times, these 
 remain ionized for times that well exceed the H\,{\sevensize II} region 
life-time (10 Myr).}
\label{fig6}
\end{figure*}

It is not a straightforward issue to estimate the fraction of
Lyman continuum photons that leaks out of galaxies. 
The only direct observations of the Lyman break in nearby star-forming galaxies
are those made by the Hopkins UV telescope (HUT) (Leitherer \etal 1995). 
The analysis suggests that less than 3 percent of the 
intrinsic Lyman continuum photons escape into the intergalactic medium
(Leitherer \etal 1995). A more recent analysis of the same data 
indicates less restrictive values, between 3 percent and
57 percent (Hurwitz \etal 1997). Our models predict 
an early evolutionary phase (between blowout and the trapping of the 
ionization front within the radiative expanding shell \ie between 3 Myr 
and 5 Myr for our A1 model) during which a large amount of the UV radiation 
could leak out of a galaxy into the intergalactic medium (see 
Fig.~\ref{fig6}a). Our predictions seem to be well in line with the 
observations given that all four star-bursts observed  with HUT are much 
older than 5 Myr (Gonz\'alez-Delgado \etal 1998) \ie  a time when no escape
is expected.

The models have assumed that the production of Lyman continuum radiation 
remains constant for up to 4 $\times 10^6$ years and then, as the ionizing 
cluster ages, it drops as $t^{-5}$ causing after $10^7$ years (the expected 
life-time of the central H\,{\sevensize II} region) 
a depletion in the ionizing radiation of more than two orders of magnitude. 
An important fraction ($\epsilon$ = 0.7) of the initial UV photon 
production rate was assumed, in all cases shown in Fig.~\ref{fig6},   
to leak out of the central H\,{\sevensize II} region causing the 
ionization of the halo and of the recombining sections of the
expanding shell. Fig.~\ref{fig6} shows the  
fraction of  UV photons  able to escape the galaxy along the 
conical H\,{\sevensize II} region,
as well as the time interval during which this becomes  possible.   
Our results, the same as those from 
calculations with other values of $\epsilon$,
with the exception of the case that assumes a constant star formation rate
(Fig.~\ref{fig6}d),  
all  show that photoionization of the halo and the escape of 
UV radiation from the galaxy occurs regardless of the production of UV 
radiation (and the assumed value of $\epsilon$) and in all cases 
limited to 5 $\times 10^6$ years. Given the importance  of the
escape of the Lyman continuum radiation in relation with estimates of star 
formation
rates, a more detailed analysis of this point is in preparation.

\begin{figure}
\psfig{figure=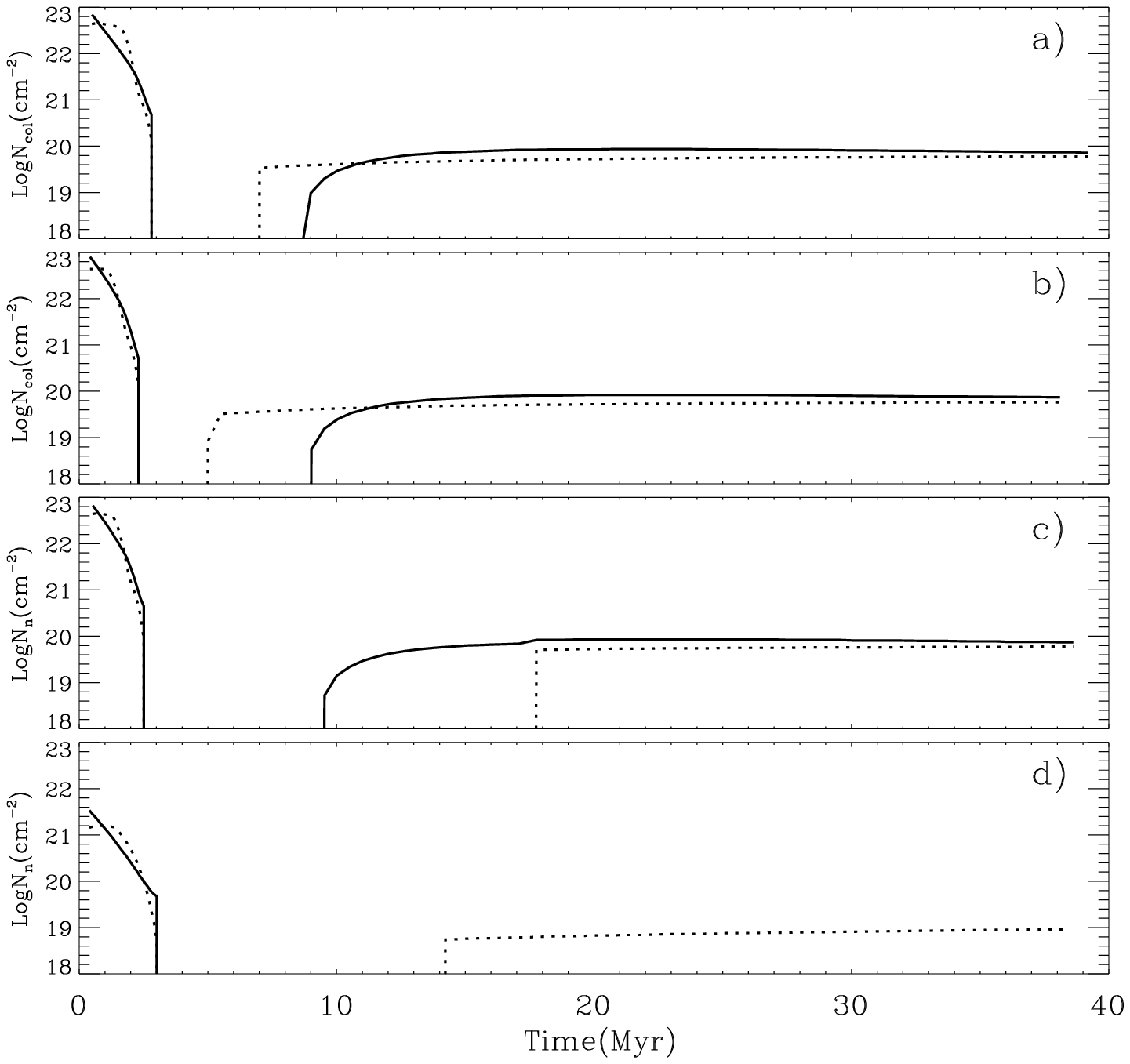,height=10cm,width=8cm,angle=0}
\vspace{1cm}
 \caption{The neutral gas column density variation along the symmetry axis for
models A1,A2, A3, and B are shown respectively in panels a -- d.
Solid lines indicate the unshocked component and the dotted lines the
column density arising from the recombining shell.}
\label{fig7}
\end{figure}

The rapid changes in ionization of the galaxies' ISM  have an impact 
on the neutral gas column density arising from the expanding shell 
and the halo, and these are central to the transport of the \la\ line emitted 
by the nuclear H\,{\sevensize II} region which otherwise will be completely 
absorbed within the extended H\,{\sevensize I} halo. If one for example 
considers the direction along the axis of symmetry, one would see that 
immediately after blowout, the ionization of the low density halo, not only 
favours the direct escape of UV photons out of the galaxy (see 
Fig.~\ref{fig6}a) but also generates a transparent region for the \la\ photons 
produced upon  recombination in the central H\,{\sevensize II} region 
(see Fig.~\ref{fig7}). 

On the other hand, the leading shock 
transition from adiabatic to radiative is to 
deplete the number of UV photons into the halo and out of the galaxy
and eventually to cause the trapping of the ionization front 
within the expanding shell. This situation favours the development of
 an expanding H\,{\sevensize I} layer and leads to
the appropriate conditions (column density and velocity) for the 
appearance of a blue shifted \la\ absorption feature. 
In model A1, full ionization of the halo is completed before 3 Myr, and
from then onwards and until the expanding shell becomes
radiative and traps the ionization front, an observer looking at the
central H\,{\sevensize II} region within the ionized halo cone should see the 
full \la\ line in emission. The trapping of the ionization front occurs 
at $t$ \si  5 Myrs, 
when the central starburst is still producing the sufficient number of 
UV photons to maintain the ionization of the central H\,{\sevensize II} region.
This indicates a high probability to observe a blue shifted \la\
absorption feature 
in intermediate mass galaxies. Recombination in the halo 
starts after 9 Myr, rapidly enhancing the column density of neutral 
material and with it the full absorption of the \la\ line produced in 
the central H\,{\sevensize II} region, irrespective of the line of sight into 
the galaxy. Panels a, b and c in Fig.~\ref{fig7} show the run of the 
H\,{\sevensize I} column density along the axis of symmetry for the standard 
model A with three different metalicities (models A1-A3).  
The change in the assumed ISM metal content simply speeds (or delays) 
the onset of radiative cooling in the expanding shell, and with it the 
trapping of the ionization front 
within the shell. Consequently, superbubbles evolving in more metal rich 
galaxies (as in model A2) 
will have a shorter lasting phase in which to show \la\ in emission,
and a correspondingly longer phase causing a blue-shifted \la\
absorption feature. The converse also holds for more metal poor galaxies 
(as in model A3).   

Despite the similar evolution of the shell  expansion velocity, which in all 
cases shows the same behaviour (at first a continuous deceleration, followed 
by a rapid acceleration phase after blowout, and a strong deceleration as the 
bubble ploughs through the extended H\,{\sevensize I} halo) 
the evolution of superbubbles in other galaxies departs in a variety of
ways from our standard case. The fact that remnants of powerful starbursts 
in massive galaxies, with steep density gradients, retain high expansion 
speeds (of several hundreds of km s$^{-1}$) along the polar sectors of 
their superbubbles for longer periods than their counterparts in less massive
galaxies (see Fig.~\ref{fig4}) is to be kept in mind. The temperature acquired 
by the shocked gas being directly related to the shock speed 
($T_{shock} = 1.4 \times 10^7(v_{shock}/1000$\kms)$^2$)
implies that these shells are to become radiative at a time that well exceeds 
both the H\,{\sevensize II} region life-time (\si 10Myr) and the 5 $\times 
10^7$ years span during which the 
starburst deposits its mechanical energy. Therefore, these galaxies 
are not likely to produce a blue shiffted \la\ absorption feature. 

Low mass galaxies appear at first glance as the best candidates to lose their
starburst processed metals, or even their ISM (Fig.~\ref{fig1}b). However, 
these systems also experience a rapid shock deceleration as their superbubbles 
evolve into the galaxy halos, leading as in the standard case A to speeds well 
below the galaxy escape velocity (see Fig.~\ref{fig4}b). Also, as shown in 
Paper I, if the mass of the galaxy is initially further centrally condensed, 
then the larger densities and stronger cooling imply an amount of energy 
much larger than what is observed (see Marlowe \etal 1995) for the superbubble 
to reach the edge of the galaxy. On the other hand, massive disk galaxies 
with a powerful nuclear starburst generate giant superbubbles able to reach 
supersonically the outskirts of the galaxy, dumping their processed matter 
into the intergalactic medium. This happens both in the coeval starburst 
model (C1) and in the constant star formation rate one (C2) 
at a time that well surpasses 40Myr.

Fig.~\ref{fig5}d--f show a cross-sectional view of the superbubble and the 
conical H\,{\sevensize II} region developed after 15Myr of evolution for cases 
B1, C1 and C2, respectively. The low density halos in these galaxies 
inhibit recombination for times that well surpase the H\,{\sevensize II} 
region life-time. These panels are to be compared with Fig.~\ref{fig5}c 
where the recombination of the halo matter 
(after 15Myr of evolution) has led to a broad shell of recombined matter ahead 
of the evolving superbubble in an intermediate mass galaxy (case A1).   

Strong radiative cooling within the expanding shell promotes 
its transition from adiabatic to radiative and this -- if it occurs 
before 5 $\times 10^6$yr (see Fig.~\ref{fig6}) -- has an 
important impact on the number of UV photons leaking into the halo and out 
of the galaxy. In both, the low and high mass systems here considered, 
the radiative shell forms much later than in the intermediate mass galaxies, 
long after the massive stars have moved off the main sequence.

Despite the large value of $\epsilon$ = 0.7 used in all 
calculations presented here, in all models with a coeval 
starburst the leakage of UV photons into the halo and 
possibly out of the galaxy is restricted to 5Myr of evolution by the 
boundary condition imposed by stellar evolution ($F_{UV} \sim t^{-5}$). The 
drastic drop in the photon production rate after the most 
massive stars begin to move off the main sequence,
inhibits UV photons to even leak out of the central H\,{\sevensize II} 
egion, and thus clearly calculations with smaller values 
of $\epsilon$ display the same upper limit
(see Fig.~\ref{fig6}a--c). The situation is different in the 
constant star formation rate (case C2) where the number of UV 
photons escaping the galaxy remains 
almost constant throughout the evolution. This is however 
also a result of the long recombination time in the low 
density halo which makes it transparent to the starburst UV flux.   

\section{Discussion} 

\begin{figure*}
\centerline{\psfig{figure=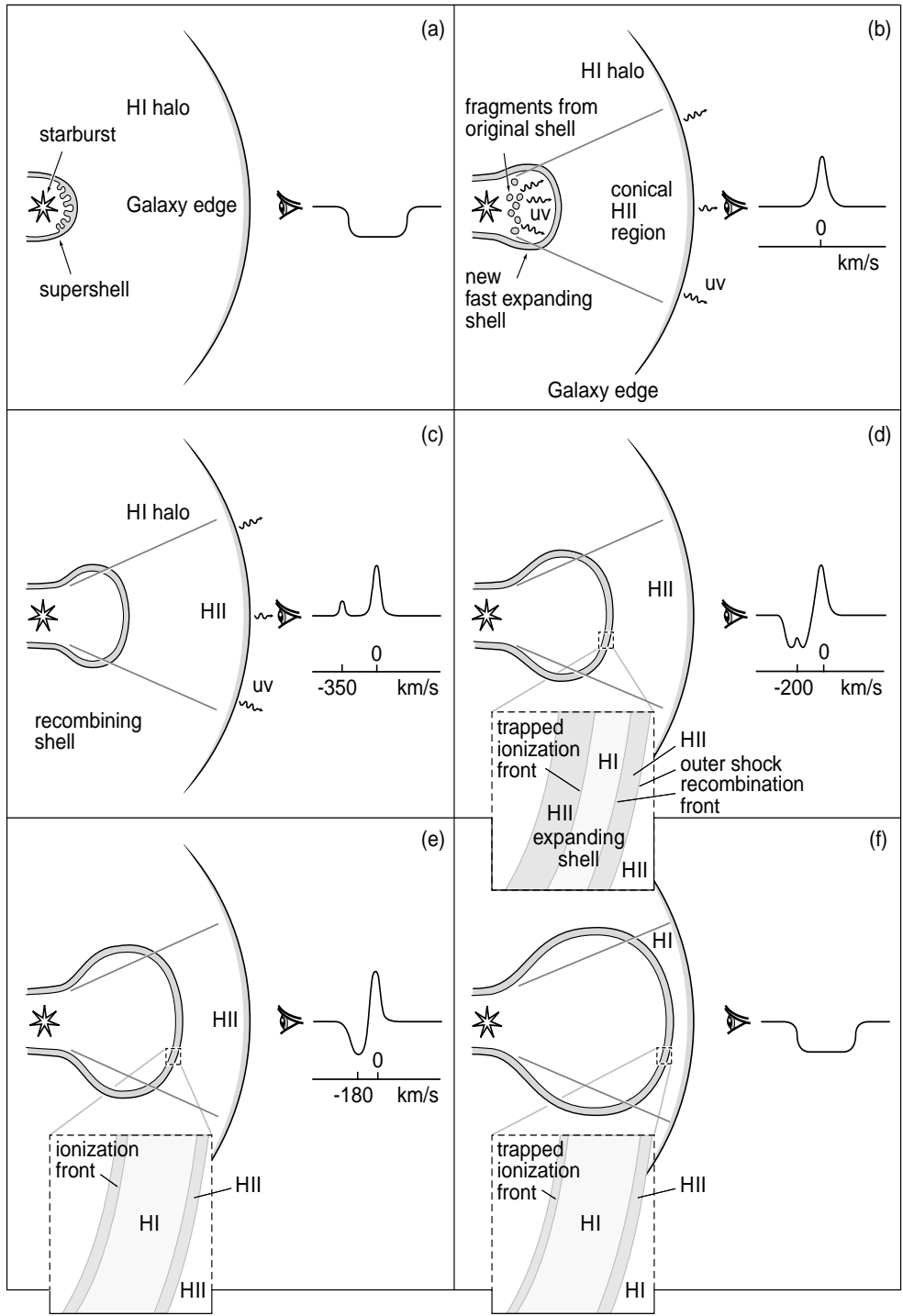,height=19cm,width=13cm,angle=0}}
\caption{The evolution of superbubbles and the \la\ emission profile expected 
from star-forming galaxies. The panels present the sequence of changes in the  
structure of the ISM produced by the evolution of a  massive starburst. On the 
right hand-side of each panel, the \la\ profile expected is shown and the 
scale corresponds to model A1 at times {\bf (a)}$\le$ 1.5Myr, 
{\bf (b)}$\sim$2.5Myr, {\bf (c)}$\sim$4Myr, {\bf (d)}$\sim$5Myr, 
{\bf (e)}$\sim$6Myr and {\bf (f)}$>$8Myr. The 
sequence starts with a giant superbubble which is about to burst into the 
extended H\,{\sevensize I} halo of a galaxy, as its shell of swept up matter 
becomes Rayleigh - Taylor unstable {\bf (a)}. The instability leads to the 
rupture of the shell and with it to the exit of the thermalized stellar ejecta 
which then freely streams and promotes the development of a new shock and a 
new shell of swept up halo material {\bf (b)}. At the same time, the stellar 
$uv$ radiation is able to leak out into the halo, and even beyond the galaxy 
outer edge, leaving behind an extended, low density,  conical 
H\,{\sevensize II} region. Recombinations in the shell of swept up halo matter 
should become increasingly more important as a function of time, causing a 
blueshifted \la\ emission and a depletion in the number of ionizing photons 
escaping the galaxy. The latter will be totally depleted once the ionization 
front is trapped by the recombining shell {\bf (c and d)}. The last stages in 
the evolution enhance the column density of H\,{\sevensize I} in the 
recombining shell leading to a P-Cygni \la\ profile and eventually to a full 
saturated absorption {\bf (e and f)}.}
\label{fig8}
\end{figure*}

Our numerical calculations of superbubbles evolving in gas-rich galaxies, 
accounting for the mechanical energy deposition as well as for the ionization 
produced by the massive members of a starburst, have led to a more complete 
picture of the impact that massive star formation may have on a galaxy ISM. 
We have centered our attention on the time evolution of superbubbles to 
evaluate the possible burst of processed matter into the intergalactic medium 
and found that this is only likely for powerful nuclear bursts in disk 
galaxies. Small and intermediate mass galaxies host superbubbles unable to 
reach their galaxy outskirts. Furthermore, one should note that the first 
10Myr of evolution, \ie the H\,{\sevensize II}
region life-time, is only a small fraction of the time 
required for superbubbles to grow and reach their largest 
dimensions, and thus studies in optical wavelengths can only 
address the beginning of the evolution and the 
possible burst into the intergalactic space occurs at much latter times.
We have also looked at the extended conical H\,{\sevensize II} region 
generated by the escape of UV photons from the central starburst once blowout 
of the superbubble into the extended H\,{\sevensize I} halo takes place. The 
early blowout into the 
extended H\,{\sevensize I} halo is promoted by the steep density drop away from
the galaxy plane, causing the burst of the thermalized ejected 
matter as well as the leak of UV radiation into the extended halo
(see Fig.~\ref{fig8}a, b). 
The evolution of the H\,{\sevensize II} region is central to the transport 
and detection of \la\ emission from the central H\,{\sevensize II} region. 
For example, once recombination starts in the fast expanding shell, 
it will cause a correspondingly blueshifted \la\ emission, as depicted 
in Fig.~\ref{fig8}c. Once the shell presents a large column density 
($\sim$ 10$^{19}$ cm$^{-2}$), as it grows to dimensions of a few kpc, it will 
trap the ionization front. This is promoted by the large shell densities and
the geometrical dilution of the ionizing radiation. Note that from then 
onwards, recombinations in the shell  will inhibit the 
further escape of ionizing photons from the galaxy (compare Fig.~\ref{fig8}b, 
c, and d). The trapping of the ionization front,  makes the shell acquire a 
multiple structure with a photoionized inner edge, a steadily growing central 
zone of H\,{\sevensize I}, and an outer collisionally ionized sector where 
the recently shocked ionized halo gas is steadily incorporated. The growth 
of the central layer eventually will cause the sufficient scattering and 
absorption of the \la\ photons emitted by the central H\,{\sevensize II} 
region, leading to a blueshifted \la\ absorption. 
Note also that for as long as recombinations continue to occur at the 
leading edge of the shell, a blueshifted \la\ in emission will appear 
superposed on the blueshifted absorption feature (see Fig.~\ref{fig8}d). 
Recombination at the 
leading edge of the shell will become steadily less frequent, depleting the 
blueshifted \la\ in emission. This is due when the shell and its 
leading shock move into 
the outer less dense regions of the halo, and the shell recombination time, 
despite the compression at the shock, becomes larger than the dynamical time. 
At this stage, an observer looking along the conical H\,{\sevensize II} region
will detect a P-Cygni-like \la\ line profile as shown in Fig.~\ref{fig8}e. 

The geometrical dilution of the UV flux will begin to make an impact as the 
superbubble grows large. This, and the drop in the UV photon production rate 
caused by the death of the most massive stars after  
$t$ = $t_{ms}$, will enhance the
column density of neutral material in the central zone of the 
recombined shell to eventually cause the full saturated 
absorption of the \la\ line (see Fig.~\ref{fig8}f). Full saturated 
absorption has usually been accounted for by the large column density of 
the extended H\,{\sevensize I} envelope of these galaxies and thus, as in all 
models, many different orientations will suit the observations. In our
scenario however, also when observing within the solid angle defined by the 
conical H\,{\sevensize II} region formed after blowout, such  broad 
absorptions could arise well after blowout, once a large column density of 
shocked neutral material ($N$ $ \geq$ 
10$^{20}$ cm$^{-2}$) has formed ahead of the trapped ionization front.

The solid angle subtended by the ionized cone will be a rapidly changing 
function of time, particularly during the early stages of evolution, 
immediately after blowout. However, numerical experiments and observations
(see Paper I, and Tenorio-Tagle \& Mu\~noz-Tu\~non 1997, 1998 
and references therein) restrict this to a 
maximum value of about 70\deg , with the walls of the superbubble
near the galaxy plane inhibiting its further growth. 

P-Cygni \la\ profiles are predicted when observing along the angle subtended 
by the conical H\,{\sevensize II} region but only once 
the ionization front is trapped by the sector of the superbubble shell
evolving into the extended halo. This will produce the fast moving 
layer of H\,{\sevensize I} at the 
superbubble shell, here thought to be responsible for 
the partial absorption observed in sources such as Haro 2, ESO 400-G043 (which
probably exhibits a secondary blueshifted \la\ emission) and
 ESO 350-IG038 (Kunth  \etal 1998). In the latter case however, the profile
is not typical of a clear P-Cygni profile. Instead the underlying  damped \la\
 absorption extends beyond the red of the line emission. 
Damped \la\ absorption is shown by several galaxies. We note that 
these objects are all gas rich dwarf galaxies whereas in most cases but 
Haro 2, the galaxies that exhibit \la\ in emission or with a P-Cygni profile,
are on the higher luminosity side of the distribution ($M$ $\leq$ $-18$).
Pure \la\ emission is observed in C0840+1201 and T1247-232 (Terlevich 
\etal 1993; IUE)  or T1214-277 (Thuan \& Izotov 1997; HST). Such a line 
implies no absorption and thus no H\,{\sevensize I} gas between the starburst 
H\,{\sevensize II} region and the observer, as when observing the central 
H\,{\sevensize II} region after the superbubble blowout, within the conical 
H\,{\sevensize II} region sector carved  in the extended H\,{\sevensize I} 
halo. 

The main implication of our calculations is that it is the feedback from the 
massive stars what -- through ionization  and the evolution of superbubbles --
leads to  the large variety of \la\ emission profiles. The escape of \la\ 
photons depends sensitively on the column density of the neutral gas and dust 
following the suggestion  that the attenuation by dust is enhanced by 
scattering with hydrogen atoms. Note that apart from 
the observed profiles: \la\ in emission, P-Cygni-like profiles and 
full saturated absorption (as in Fig.~\ref{fig8} a, b, e and f) the scenario 
predicts also secondary blueshifted \la\ emission profiles emanating 
from the rapidly expanding and recombining shell (see 
Fig.~\ref{fig8}c and d). If massive star formation leads also to networks of
shells such as those observed in 30 Dor (Chu \& Kennicutt 1992) one 
should also expect a forest of \la\ in emission arising from 
recombinations in the various expanding shells in the network.

\section*{Acknowledgments}

The authors wish to thank the Guillermo Haro Advanced program of Astrophysics 
at INAOE, during which this work was completed. 
GTT, ET and RT acknowledge support from an EC -- EARA  -- grant and CNRS 
during visits to IAP where part of this work was accomplished. 
SAS acknowledges support from the Royal Society grant for joint projects
with the former Soviet Union, and the RGO and the IoA in 
Cambridge for partial financial support. SAS also thanks the INAOE -- while ET 
thanks the RGO and IoA -- for hospitality and partial financial support.
We thank Stephan Charlot and Max Pettini for  comments and suggestions and 
Richard  Sword from the IoA for his help with the drawing of Fig.~\ref{fig8}.

\label{lastpage}

\end{document}